\documentclass[12pt,a4paper]{article}

\usepackage{SGpreprint}
\usepackage{url,lastpage}

\nologohead{Emeterio Navarro, Rub\'en
Cantero, Jo\~ao  Rodrigues, Frank Schweitzer\\
Investments in Random Environments \\
\textsl{Physica A} vol. 387, no. 8-9 (2008), pp. 2035-2046}

\usepackage{epsfig}
\usepackage{amsmath}
\usepackage[sort&compress]{natbib}

\begin{document}

\begin{center}

   \textbf{\Large Investments in Random Environments}\\[5mm]

\textbf{Jes\'us Emeterio Navarro-Barrientos$^{1}$, Rub\'en
Cantero-\'Alvarez$^{2}$, \\ Jo\~ao F. Matias Rodrigues$^{3}$,
Frank Schweitzer$^{3,\star}$}

\begin{quote}
  \begin{itemize}
  \item[$^{1}$]{Institute for Informatics, Humboldt University, Unter den
      Linden 6, 10099 Berlin, Germany}

  \item[$^{2}$]{Institute for Applied Mathematics, University of Bonn,
      Wegelerstr. 10, 53115 Bonn, Germany}
  \item[$^{3}$] {Chair of Systems Design, ETH Zurich, Kreuzplatz 5, 8032
      Zurich, Switzerland}
  \item[$^{\star}$] Corresponding author,
    \url{fschweitzer@ethz.ch}, \url{http://www.sg.ethz.ch}
\end{itemize}
\end{quote}

\end{center}

\begin{abstract}
  We present analytical investigations of a multiplicative stochastic
  process that models a simple investor dynamics in a random environment.
  The dynamics of the investor's budget, $x(t)$, depends on the  
  stochasticity of the return on investment, $r(t)$, for which different
  model assumptions are discussed. The fat-tail distribution of the  
  budget is investigated and compared with theoretical predictions.  We  
  are mainly interested in the most probable value $x_{\mathrm{mp}}$ of
  the budget that reaches a constant value over time. Based on an
  analytical investigation of the dynamics, we are able to predict  
  $x_{\mathrm{mp}}^{\mathrm{stat}}$. We find a scaling law that relates  
  the most probable value to the characteristic parameters describing the
  stochastic process. Our analytical results are confirmed by stochastic
  computer simulations that show a very good agreement with the  
  predictions.
\end{abstract}


\newcommand{\mean}[1]{\left\langle #1 \right\rangle}
\newcommand{\abs}[1]{\left| #1 \right|}
\section{Introduction} 
\label{sec:Introduction}  

Multiplicative stochastic processes denote type of dynamics where a
variable $x(t)$ changes its value in time due to a stochastic term
$\lambda(t)$:
\begin{equation}
  \label{eq:basic}
  x(t+1)=\lambda(t)\,x(t)
\end{equation}
$\lambda(t)$ may describe different stochastic processes, such as
Gaussian white noise, uniform random distributions, GARCH or ARCH
processes, which are explained later in this paper. Despite its great
simplicity, the dynamics of eqn. (\ref{eq:basic}) gained much attention
in different fields of applications. It was, for example, used already in
1931 by Gibrat to describe the annual growth of companies -- an idea
extended in different works by economists \citep{sutton97, gabaix99a} and
econophysicists \citep{aoyama03a}.  Several theoretical aspects of
stochastic processes with multiplicative noise have been focus of recent
research in physics \citep{Redner90, Levy-Solomon96,
  Takayasu-Sato97,Sornette-Cont97, Sornette98}.  Moreover, by taking into
account additional couplings between these processes, the approach has
been extended towards a generalized Lotka-Volterra model
\citep{malcai,solomon02} which is referred to later in this paper.

The importance of multiplicative stochastic processes was also reflected
by several mathematical investigations \citep{Vervaat79C, horst01,
  horst04}. In the so-called Kesten process \citep{kesten73} -- from
which we depart in this paper -- the dynamics of eqn.  (\ref{eq:basic})
was extended by a second \emph{additive} stochastic term $a(t)$
independent of $\lambda(t)$
\begin{equation}
  \label{eq:basic2}
   x(t+1)=\lambda(t)\,x(t)+a(t)
\end{equation}
This extension has the nice feature that the stochastic process is
repelled from zero, provided some constraints on $a(t)$ are satisfied. We
will come back to this later in our paper. At this point, we rather want
to provide some arguments why we became interested in this topic and why
we think that this is relevant. 

Our investigations started with the question of how much ``intelligence''
is needed for an agent to survive in a noisy environment (see also
\citep{farmer2005ppz}). The example at hand is explained in Sect.
\ref{sec:InvestmentModel}: agents with a personal budget, $x(t)$,
participate in a simple investment scenario and face a return on their
investment, $r(t)$. Given some uncertainty of $r(t)$, they have the
choice to adjust individually the portion of the budget they wish to
invest, $q(t)$, which can be called their risk propensity. The strategy
to choose $q(t)$ may of course depend on whether the agents are able to
obtain some information about the expected return on investment, $r(t)$.
Here, different levels of internal complexity, or ``intelligence'', of
agents come into play, namely their capability to observe and to store a
history of previous returns $r(t-\tau)$, to calculate different measures
from such a time series (such as trends, moving least squares, etc.), to
detect or to forecast certain periodicities in the signal received (such
as cycles in the market) \citep{Navarro,NavarroSchweitzer03}. At this
point, a wealth of different models, assumptions, suggestions,
speculations etc. about the agent's behavior comes into play from various
fields (financial theory \citep{markowitz91, LeBaron00, farmer2005ppz,
  Raberto-Cincotti03}, economics \citep{lux2000vcf,Takayuki-Kurihara04,
  Carpenter02, Klos-Nooteboom, Parkes-Huberman01, Gode-Sunder93},
behavioral sciences \citep{Takahashi-Terano03}, physics \citep{maslov98,
  Mantegna-Stanley-2000, Daniels-Farmer03}, computer sciences
\citep{Lawrenz-Westerhoff03, Drake-Marks02, Kassicieh-Paez-Vora98}). All
these rather complex assumtions eventually lead to their specific outcome
of the game, and can hardly be compared.  So, before embarking on some
more refined modeling assumptions about agents behavior, we had to ask
ourselves what could be the reference case in this simple investment
scenario, to which later all advanced simulations can be compared. In
other words: what would be the dynamics of that investment process
\emph{without} all these ``intelligent'' assumptions? It turns out that
this baseline case is exactly given by the dynamics of a multiplicative
stochastic process with an additive term as described by eqn.
(\ref{eq:basic2}). Instead of a rather complex strategy $q(t)$, we simply
assume $q_{0}=\mathrm{const.}$ as the reference case, instead of an
unknown additional influx $a(t)$, we take a constant, but positive
``income'' $a$, and instead of specific economic assumptions about market
fluctuations we choose $r(t)$ from four different stochastic processes,
which are simple, but analytically tractable. This way, in Sect.
\ref{sec:InvestmentModel} we arrive at the basic dynamics of eqn.
(\ref{eq:budgetmap}), which is equivalent to our starting equation
(\ref{eq:basic2}).

The aim of our paper is (i) to elucidate the dynamics of eqn.
(\ref{eq:basic2}) for some specific settings of $\lambda(t)$ and $a(t)$
by means of some computer simulations, and (ii) to investigate its
stationary properties by means of analytical treatment. This will reveal
some scaling between the most probable value of $x(t)$ and the parameters
describing the stochastic processes. Before we continue in this
direction, we want to shortly summarize some previous theoretical
investigations of eqn. (\ref{eq:basic2}), in order to refer to it later.

The dynamics of eqn. (\ref{eq:basic2}) was treated by
\citet{Sornette-Cont97} as
\begin{equation}
  \label{eq:diff_map}
  \frac{x(t+1)-x(t)}{x(t)}=\frac{a(t)}{x(t)}+\lambda(t)-1
\end{equation}
If the finite difference ${\left(x(t+1)-x(t)\right)}/{x(t)}$ is approximated
by $(d\log x/dt)$, the following overdamped Langevin equation for
$w=\log{x}$ can be obtained:
\begin{equation}
  \label{eq:langevin_map}
  \frac{dw}{dt}=a(t)e^{-w}-\abs{\upsilon}+\eta(t)
\end{equation}
with:
\begin{equation}
  \label{eq:langevin2_map}
  \upsilon=\mean{\lambda}-1\simeq \mean{\log \lambda} \;\quad
  \mean{\eta^2} =\mean{\lambda^2}-\mean{\lambda}^2
\end{equation}
The first term on the r.h.s. of eqn. (\ref{eq:langevin_map}) describes an
effective repulsion of $x$ from zero, while the second term,
$-\abs{\upsilon}$, describes the drift towards zero.  The third term,
$\eta(t)$, expresses the stochastic influences resulting from
$\lambda(t)$.

Going over from the single stochastic realizations of $\omega(t)$ to the
probability density $P(\omega,t)$, it was shown in
\citep{Sornette-Cont97} that the following Fokker-Planck
equation\footnote{Note that this equation is different from the original
  one in \citep{Sornette-Cont97} by a factor 2 in the diffusion term
  which later affects the definition of $\mu$ in eqn. (\ref{eq:mu})} can
be derived
\begin{eqnarray}
  \label{eq:fokker_map}
  \frac{\partial P(w,t)}{\partial t} & = & a(t)\,e^{-w}\,P(w,t)
  -\left(\mean{\log{\lambda}}+a(t)\,e^{-w}\right)\frac{\partial P(w,t)}
{\partial w} \nonumber \\
  & & 
  +\left(\mean{\log{(\lambda)^2}}-\mean{\log{\lambda}}^2\right)\,
\frac{1}{2}\frac{\partial^2
  P(w,t)}{\partial w^2} 
\end{eqnarray}
In accordance with eqn. (\ref{eq:langevin_map}), the first term of eqn.
(\ref{eq:fokker_map}) describes the decay on $w$, the second term
indicates the drift of the process and the third is a diffusion term with
the diffusion constant:
\begin{equation}
\label{eq:diff}
  D=\mean{(\log{\lambda})^2}-\mean{\log{\lambda}}^2
\end{equation}

We recall that \emph{without} the additive stochastic force, i.e.
$a(t)\equiv 0$, eqn. (\ref{eq:fokker_map}) results in a simple
Fokker-Planck equation for $x(t)$, related to the stochastic eqn.
(\ref{eq:basic}), with the \emph{log-normal distribution} as limit
distribution \citep{Redner90,Sornette-Cont97}:
\begin{equation}
  \label{eq:lognormal_mp}
  P(x(t))=\frac{1}{\sqrt{\pi\,D\,t}}\,\frac{1}{x(t)}\,
\exp{\left\{-\frac{1}{D\,t}
\big[\log{x(t)}-\mean{\log{\lambda}}\,t\big]^2\right\}}
\end{equation}

Considering, however, an additive stochastic force $a(t)\not\equiv 0$, it
was already noted in \citep{kesten73} that for \emph{large} $x$ instead
of a log-normal distribution now a power-law distribution results,
provided that $\mean{\log \lambda}<0$:
\begin{equation}
 \label{eq:powerlaw}
 P(x) \propto x^{-(1+\mu)}
\end{equation}
The exponent $\mu$ satisfies the conditions
\begin{equation}
  \label{eq:mu}
\mean{\lambda^{\mu}}=1  \;;\quad 
\mu=\frac{2\abs{\mean{\log{\lambda}}}}{
\mean{(\log{\lambda})^2}-\mean{\log{\lambda}}^2} 
=\frac{2\abs{v}}{D}
\end{equation}
It was shown \citep{Sornette-Cont97} that such power-laws result under
quite general assumptions about multiplicative stochastic processes with
repulsion at the origin which generalizes previous results by
\cite{Levy-Solomon96}. We note that an economically motivated entry/exit
dynamics \citep{simon-bunoni,Blank-Solomon00, Richiardi04} as a boundary
condition of the multiplicative stochastic process also leads to power
distributions.

In the more general case of eqn. (\ref{eq:fokker_map}), we will in Sect.
\ref{sec:stationary} provide a stationary solution for $P(\omega)$ at
least for the case of a constant $a$. We will further use this solution
to derive a result for the most probable value of the multiplicative
stochastic process. In order to test the analytical predictions, we will
compare them with stochastic computer simulations, which are described in
the following section.

\section{Computer simulations for different random environments}

\subsection{Investment model}
\label{sec:InvestmentModel}

Instead of solving the stochastic differential equation
(\ref{eq:fokker_map}) numerically, in this section we focus on the
individual realizations of the stochastic process as described by eqn.
(\ref{eq:basic2}). Hence, we have to specify the stochastic processes
behind the variables $\lambda(t)$ and $a(t)$, which is done in the
following.  As it became clear from the previous discussion, the additive
stochastic term acts as a repulsion of the dynamics from zero, i.e.
without $a(t)$ after some time every stochastic realization of $x(t)$
will reach the value zero, i.e., the exit for the process. So, the
meaning of $a(t)$ is simply to keep the dynamics ``alive'' by preventing
it from reaching zero.  To simplify the dynamics, instead of a
time-dependent value $a(t)$ we simply choose a small, but constant
positive value of $a$, which results in the same effect.

For the multiplicative stochastic term the following assumption was made: 
\begin{equation}
  \label{eq:lambda_budget}
  \lambda(t)=1+r(t)\,q(t)
\end{equation}
In order to give the stochastic dynamics of eqs. (\ref{eq:basic2}),
(\ref{eq:lambda_budget}) some interpretation, let us consider the
following: $x(t)$ shall represent the budget (wealth, liquidity) of an
agent, who intents to invest some portion $0 \leq q(t)\leq 1$ of its
budget into a market \citep{NavarroSchweitzer03,Navarro}.
Depending on the market performance, this investment may result in a loss
or a gain.  Hence, there exists a \emph{return on investment}, RoI,
$r(t)$, which is either positive or negative.  It should be noted that
the RoI has to satisfy $r(t)>-1$ as a lower boundary, as investors can
never loose more than what they have invested. Because there is no limit
for potential profits, in principle there is no upper boundary for
$r(t)$.

In a real investment scenario, the RoI can be determined from a time
series of a real market that is influenced by several thousands bids and
asks. This would involve to model the market dynamics explicitely. In the
spirit of other investigations in economics and econophysics
\citep{aoyama03a} we have chosen instead to model the market return
$r(t)$ by means of different stochastic processes, i.e. to keep its
dynamics independent of the investment of the agent. Further, we have
bound the RoI to values between -1 and +1. The upper boundary was chosen
to obtain a mean of zero for $r(t)$, which allows to better understand
the basic dynamics.  The following distributions for $r(t)$ are
discussed:
\begin{enumerate}
\item $B\{-1,1\}$, a binary stochastic switch between only two states, -1
  and +1
\item $U(-1,1)$, a uniform distribution, where every possible value
  between -1 and +1 has the same probability to be chosen
\item $N(0,0.1)$, a normal distribution with mean $0$ and standard
  deviation 0.1, i.e. values close to the mean are more likely to be
  picked
\item ARCH(1), a distribution, where possible values between -1 and +1
  are chosen by considering some correlations between times $t$ and $t-1$
  (autoregressive conditional heteroskedasticity) \citep{Bera-Higgins93,
    Engle82}.
\end{enumerate}
These stochastic processes were not chosen in the first place because
they capture \emph{real} economic processes e.g. on financial markets,
but because they later allow for analytical calculations of some
properties of the probability distributions, as shown in Sect.
\ref{sec:scaling}. However, we note that the four types cover different
degrees of complexity in the stochastic process, as their spectrum of
values (discrete, continuous) and their range of values (rather broad for
the uniform distribution, but narrowed and centered for the normal
distribution) vary and even correlations between different time steps are
taken into account (ARCH(1)).

The investment dynamics is then described by the multiplicative
stochastic process
\begin{equation}
  \label{eq:budgetmap}
  x(t+1)=x(t)\,\left[1+r(t)\,q(t)\right]+a
\end{equation}
where the positive constant $a$ prevents the investor from going
bankrupt.

The challenge for an agent in this very simple investment model is then
to adjust its portion of the budget to be invested, $q(t)$, given some
(observed) market returns $r(t)$.  Low values of $q$ refer to a risk
averse investment strategies, while higher values may be more risky, but
also more rewarding with respect to the budget. In \citep{Navarro}
we have discussed different scenarios for agents to adjust their risk
propensity, $q(t)$. In this paper we focus more on the stochastic
dynamics and therefore just choose $q(t)=q_{0}$, which is a small but
constant value.  This can serve as a reference case for more complex
investment strategies \citep{Navarro}.

At this point, we wish to note that the dynamics proposed in eqn.
(\ref{eq:budgetmap}) is much simpler than the comparable starting
equation in the generalized Lotka-Volterra model \citep{malcai,solomon02}
mentioned in Sect. \ref{sec:Introduction},
\begin{equation}
x_{i}(t+1)=x_{i}(t)\,\left[1+\lambda(t)\right]+a\bar{x}(t)-C x_{i}(t)
\label{eq:LV}
\end{equation}
which assumes additional couplings between different individual
stochastic processes, $x_{i}(t)$. Instead of a small, but constant income
$a$ the term $a\bar{x}(t)$ considers a \emph{global coupling} via the
mean budget $\bar{x}(t)$ of all agents (which may be related to general
publicly funded services). Moreover, the third term $C x_{i}(t)$
describes direct interactions between different agents, as $C$ is a
function dependent on other $x_{j}$. These may account for competition
for limited resources and saturation effects in the dynamics. Even if,
after some approximations discussed in \citep{malcai}, these additional
influences may be small, they still affect the general solutions for the
underlying probability distributions as we will discuss in Sect.
\ref{sec:stationary}.  If we recast the differences between eqs.
(\ref{eq:budgetmap}), (\ref{eq:LV}) using the general framework of
multiplicative processes \citep{richmond01EPJB}
\begin{equation}
  \label{eq:GF}
  \Delta x(t)=\eta(t) G[x(t)] + F[x(t)]
\end{equation}
where $\eta(t)$ is a stochastic variable, then we have for our model,
eqn. (\ref{eq:budgetmap}) $G(x)=x$ and $F(x)=a$, whereas for
eqn. (\ref{eq:LV}) $G(x)=x$ and $F(x)=a(1-x)$ results  \citep{malcai}. 
This will eventually affect the specific exponents of the stationary
probability distributions, $P_{s}(x)$. 

\subsection{Simulation results}
\label{sec:Simulations}

Here, we present stochastic computer simulations of eqn.
(\ref{eq:budgetmap}) for different distributions of $r(t)$ as described
in the previous section. Initially, $x(0)=10$ holds for the agent's
budget, $q_{0}$ and $a$ are kept constant during each simulation.  The
distributions were realized using 10'000 agents. In the case of the time
evolution of the most probable value $x_{mp}$, the plotted values were
calculated averaging over the $x_{mp}$ obtained for distributions
resulting from 10 different realizations of the simulation.

\begin{figure}[htbp]
  \begin{center}     \includegraphics[width=6.8cm]
    {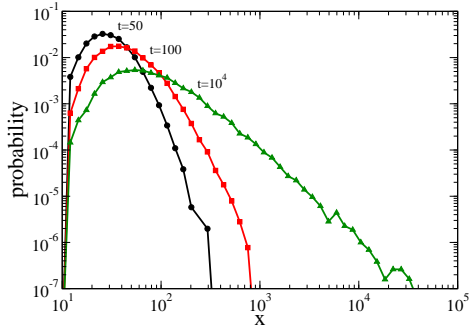}
    \caption{Investor's budget probability distribution $P(x,t)$ for
      different time steps, assuming $x(0)=10$ and $r(t)\in B\{-1,1\}$,
      $q_{0}=0.1$ and $a=0.5$. The probability was estimated from
      $N=10^{4}$ runs. }
          \label{fig:px}   \end{center}
\end{figure}

Figure \ref{fig:px} shows the distribution of the investor's budget at
different time steps. Starting from the delta distribution at $x(0)=10$,
the distribution disperses over time until it reaches a stationary state.
The number of time steps needed to reach this stationary state depends on
the initial conditions, the distribution of $r(t)$ and the additive
constant $a$. For the conditions in Figure \ref{fig:px}, this takes about
$10^4$ iterations.

\begin{figure}[htbp]
  \begin{center}    
    \includegraphics[width=6.6cm]{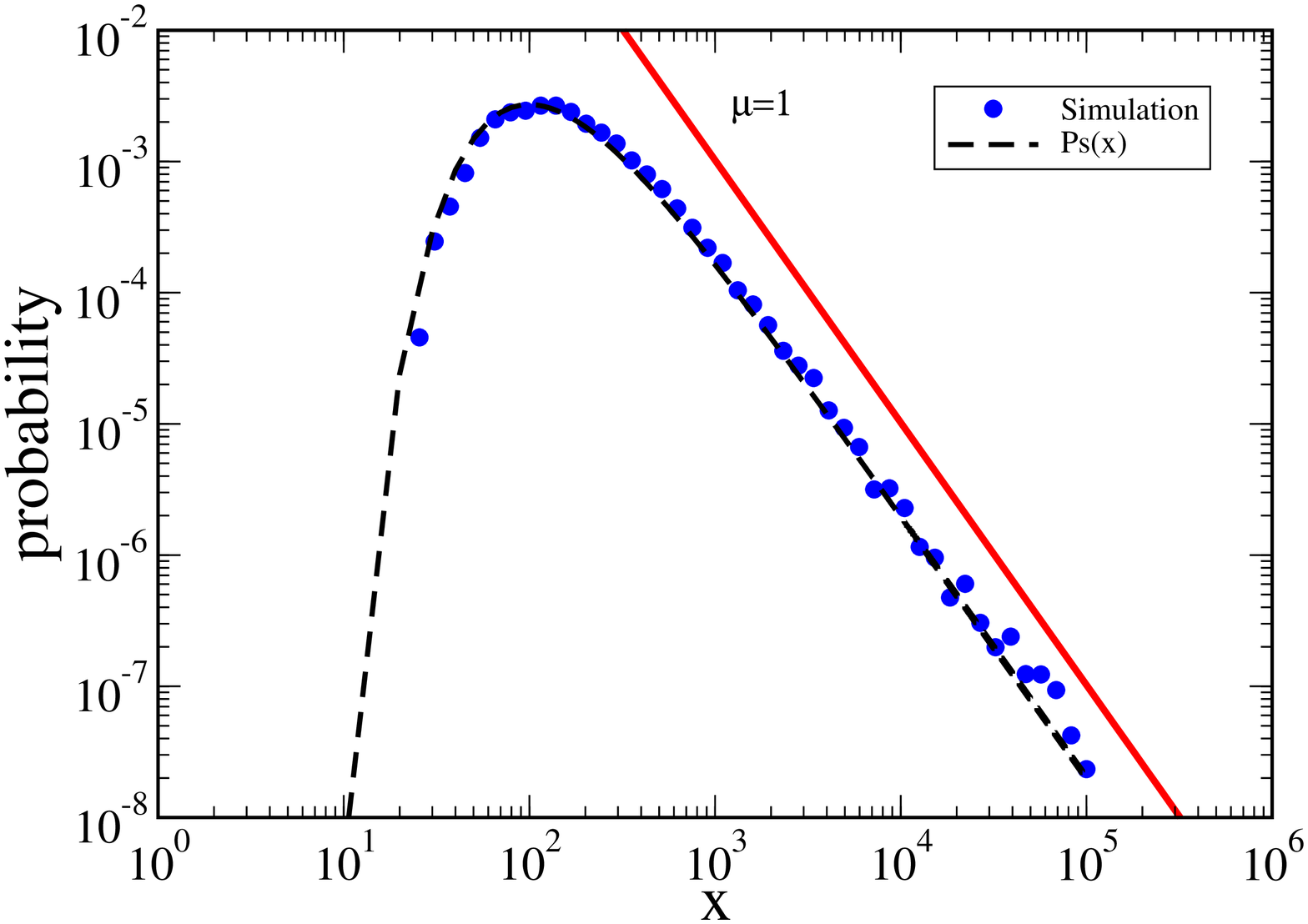}
\hfill
    \includegraphics[width=6.6cm]{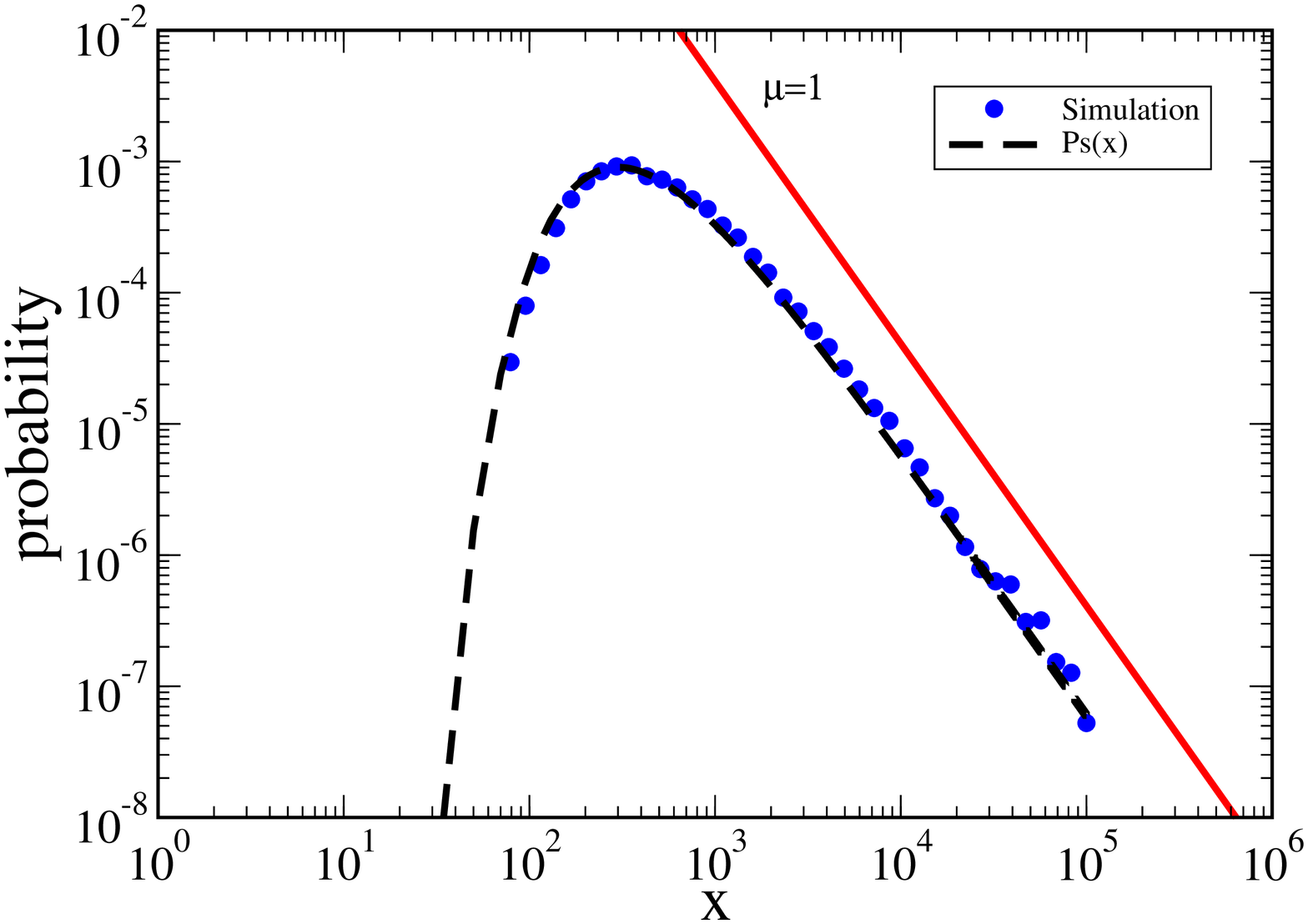} 
    \caption{Investor's budget stationary probability distribution
      $P_{s}(x)$ (estimated from frequencies after $t=10^{4}$): (left)
      Binary stochastic return distribution $r(t)=B\{-1,1\}$. (right)
      Uniform stochastic return distributions $r(t)=U(-1,1)$. In both
      cases, $x(0)=10$, $q(t)=0.1$ and $a=1$.  Data is binned in
      logarithmic intervals of the same size. The dashed lines show the
      theoretical prediction of these curves by Eqn.  (\ref{eq:psx}),
      derived later.
      \label{fig:x1}}  
  \end{center}
\end{figure}

One can clearly see that the stationary distribution is characterized by
a fat tail described by a power-law distribution, as shown in Figures
\ref{fig:x1}.  This is in accordance with previous investigations
\citep{Sornette98,Levy-Solomon96} and was already discussed in Sect.
\ref{sec:Introduction}.  We have determined the scaling exponent $\mu$ of
the power law, eqn.  \ref{eq:powerlaw}, from the simulation data for
different stochastic processes and will later compare it with our
analytical investigations.

\begin{figure}[htbp]
  \begin{center}    
    \includegraphics[width=6.6cm]{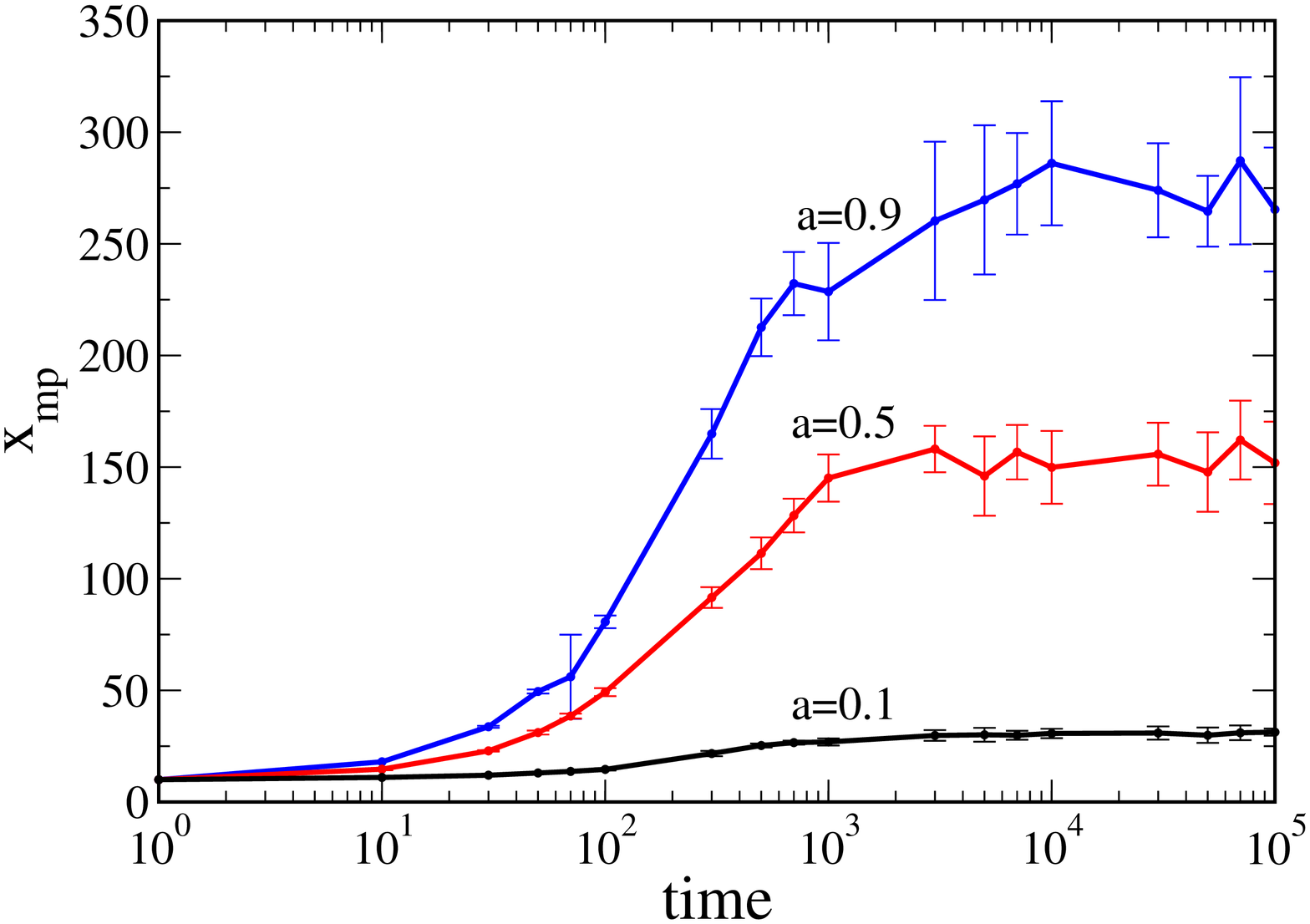}\hfill    
    \includegraphics[width=6.6cm]{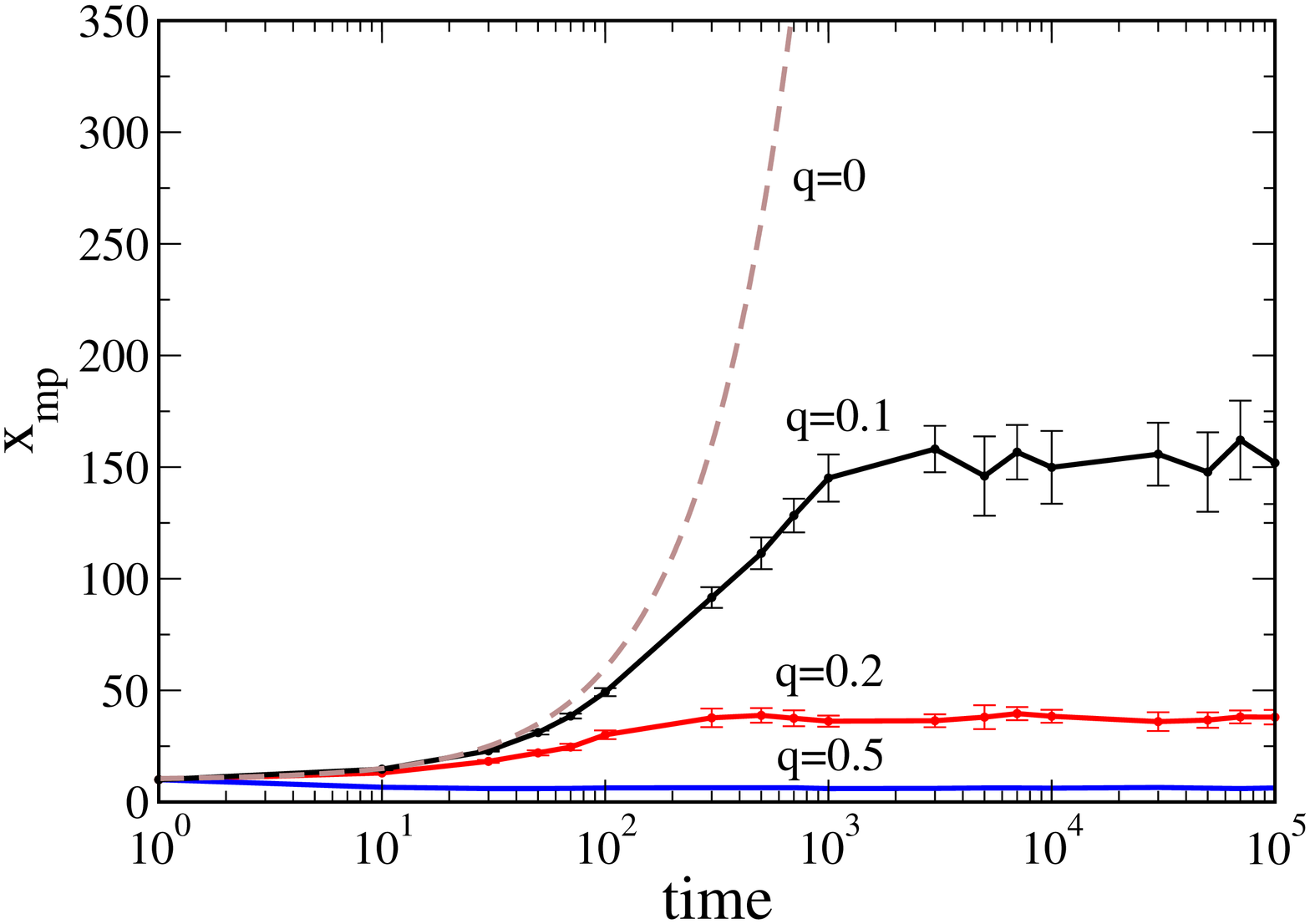}  
    \caption{Most probable budget value vs. time: (left) for $q=0.1$ and
      3 different additive constants $a$, (right) for $a=0.5$ and 3
      different constant risk propensities $q_{0}$. Other parameters and
      settings as in Figure \ref{fig:x1}(right).}  \label{fig:xmp_diffa}  
  \end{center}
\end{figure}

In the following, we are more interested in the clear maximum of the
stationary distribution which gives the \emph{most probable value},
$x_{\mathrm{mp}}$.  Figure \ref{fig:xmp_diffa} (right) shows the
evolution of $x_{\mathrm{mp}}$ over time for three different additive
terms. The most probable value increases with time until a balanced state
is reached, where the dissipation described by the negative drift towards
zero and the constant influx $a$ compensate on average.  It can be
noticed that the standard deviation of $x_{\mathrm{mp}}$ increases with
$a$.  Larger $a$ lead to larger values for the budgets, this in turn
leads to an amplification of the fluctuations due to the multiplicative
nature of the process.  Figure \ref{fig:xmp_diffa} shows the evolution of
$x_{\mathrm{mp}}$ for three different values of the risk propensity
$q_{0}$. Obviously, the larger $q$, the smaller $x_{\mathrm{mp}}$.

\begin{figure}[htbp]
  \begin{center}
 \includegraphics[width=6.8cm]{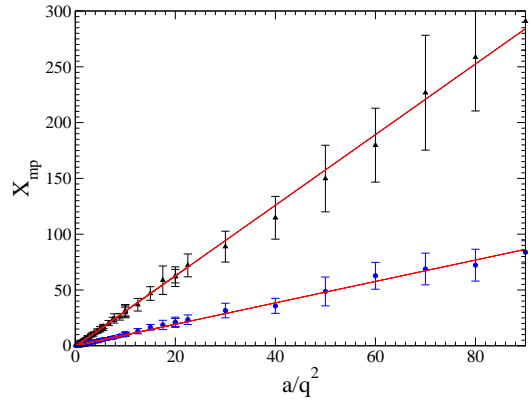}
 \caption{Most probable budget value vs. scaled variable      
   $a/q_{0}^{2}$ for the binary stochastic process $r(t)\in \{-1,1\}$ (o)
   and for the uniform stochastic process $r(t)\in U(-1,1)$
   ($\triangle$). Each set was plotted by varying $a$ and $q_0$ over the
   range of $[0.1,0.9]$ in $0.1$ increments, giving a total of 81 data
   points per combination, , averaged over 100'000 Monte-Carlo
   simulations and 10 runs. The value of the slope found for the
   numerical simulations is for the binary stochastic process
   $c=0.961\pm0.006$ and the uniform stochastic process $c=3.149\pm0.014$
 }
    \label{fig:xmp_scaling2RtU}
  \end{center}
\end{figure}

\begin{figure}[htbp]
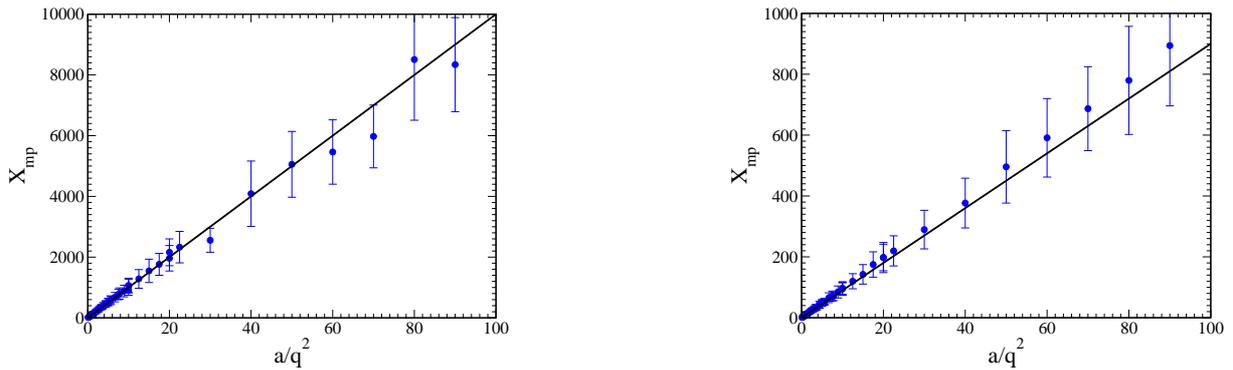
  
  \begin{center}
\includegraphics[width=6.6cm]{figures/xmpGaussianpval0_7}\hfill
\includegraphics[width=6.5cm,angle=0]{figures/xmpARCHpval0_5}
    \caption{Most probable budget value vs. scaled variable      
      $a/q_{0}^{2}$: (left) for the normal distributed stochastic process
            $r(t)\in N(0,0.1)$.  The value found for the slope of the
            numerical simulations in this process was:
            $c=102.17\pm0.701$, (right) for the ARCH(1) process
            ($\alpha_{0}=\alpha_{1}=0.1$), with a slope of
            $c=9.089\pm0.029$}
    \label{fig:xmp_scalingN}
  \end{center}
\end{figure}

Figure \ref{fig:xmp_diffa} suggest that there is a scaling between the
most probable value $x_{\mathrm{mp}}$ and the parameters characterizing
the stochastic dynamics, eqn.  (\ref{eq:budgetmap}), namely $q_{0}$ and
$a$. This scaling is investigated numerically in the following and will
later be confirmed by analytical investigations. Figures
\ref{fig:xmp_scaling2RtU}, \ref{fig:xmp_scalingN} show
$\mean{x_{\mathrm{mp}}}$ at a fixed time, $t=10^{4}$, for the four
different realizations of the stochastic process, $r(t)$, discussed in
Sect. \ref{sec:InvestmentModel}.   In all four cases, the results,
plotted against the variable $a/q_{0}^{2}$ clearly show a straight line,
which allows for the scaling:
\begin{equation}
  \label{eq:scaling}
  x_{\mathrm{mp}}=c\cdot \frac{a}{q_{0}^{2}}
\end{equation}

By means of analytical investigations of the stochastic multiplicative
dynamics, we now want to confirm the findings of the numerical
experiments. 

\section{Analytical investigations}
\label{sec:analytical}

\subsection{Stationary state}

\label{sec:stationary}

In order to derive the observed scaling law, eqn. (\ref{eq:scaling}), we
start with the Fokker-Planck eqn.~(\ref{eq:fokker_map}).\footnote{In the
  appendix, we point to the direct treatment of the stochastic dynamics,
  eqn. (\ref{eq:basic2}).} Using the
approximation of a constant value of $a$ 
and the notation of eqn. (\ref{eq:diff}) for the diffusion constant $D$,
we find for the stationary solution of eqn. (\ref{eq:fokker_map}):
\begin{equation}
\label{eq:stat1}
0=-\partial_w\left[\left(\mean{\log\lambda}+a\,e^{-w}\right)P_s(w)
-\frac{D}{2}\partial_wP_s(w)\right]
\end{equation}
This results in the  stationary solution:
\begin{equation}
\label{eq:stat2}
P_s(w)=\mathcal{N}\exp\left(\frac{2\mean{\log\lambda}\,w-2a\,e^{-w}}
{D}\right)
\end{equation}
with normalisation $\mathcal{N}$. Using $w=\log x$, the corresponding
stationary probability distribution $P_{s}(x)$ is recovered by the chain
rule:
\begin{equation}
\label{eq:statx}
P_{s}(x)=P_{s}\left(w(x)\right)\,\frac{dw}{dx} 
=\mathcal{N}\,x^{\frac{2\mean{\log\lambda}}{D}-1}\,
\exp\left(-\frac{2a}{D\,x}\right).
\end{equation}
The normalization is explicitely calculated via:
\begin{equation}
\label{eq:norm1}
1\stackrel{!}{=}\mathcal{N}\int_0^{\infty}x^{\frac{2\mean{\log\lambda}}
{D}-1}\,\exp\left(-\frac{2a}{D\,x}\right)\,dx
=\mathcal{N}
\left(\frac{D}{2a}\right)^{-\frac{2\mean{\log\lambda}}
{D}}\Gamma\left(-\frac{2\mean{\log\lambda}}
{D}\right)
\end{equation}
where $\Gamma(y)$ is the Gamma function. Redefining
\begin{equation}
\mu:=-\frac{2\mean{\log\lambda}}{D},
\label{eq:mu2}
\end{equation}
we eventually find the stationary distribution in the normalized form:
\begin{equation}
  \label{eq:psx}
  P_{s}(x)=\frac{\left(\frac{2a}{D}\right)^{\mu}}{\Gamma(\mu)}
  \,x^{-(1+\mu)}\,\exp\left(-\frac{2a}{D\,x}\right).
\end{equation}
A graphical visualization of this function can be seen in Figs.
\ref{fig:x1} together with the results obtained from the averaging of the
stochastic computer simulations.  It is worth noticing that the Gamma
function diverges for $\mu\rightarrow0$. Hence, the condition of
$\mean{\log\lambda}<0$ is indeed important for the existence of the
stationary solution. One should also notice that for large $x$ eqn.
(\ref{eq:psx}) reduces to the power-law distribution, eqn.
(\ref{eq:powerlaw}).  

We note that eqn. (\ref{eq:psx}) is a special form of the general
solution
\begin{equation}
  \label{eq:GFPS}
  P_{s}(x)=\frac{1}{G^{2}(x)}\,\exp\left(\frac{2}{D}\int^{x}
    \frac{F(x^{\prime})}{G^{2}(x^{\prime})}dx^{\prime}\right) 
\end{equation}
obtained in \citep{malcai, richmond01EPJB}. If we use $F(x)=a$ and
$G(x)=x$ in accordance with the stochastic process defined in eqn.
(\ref{eq:budgetmap}), this would lead to the (non-normalized) solution
$P_{s}(x)=x^{-2} \exp(-2a/Dx)$. This is in agreement with eqn.
(\ref{eq:psx}) because of $\mu\approx 1$ in our case. The solution
discussed in \citep{malcai} is however different from ours because
$F(x)=a(1-x)$ and $G(x)=x$ was used there, which eventually lead to the
stationary distribution $P_{s}(x)=x^{-2(1-a/D)} \exp(-2a/Dx)$, and
consequently to $\mu=1+(2a/D)$. Only in the limit of $a\ll D$ (which may
hold for the case discussed in \citep{malcai} because of e.g. $a\approx
D/4$, but is hardly satisfied in our model, because for example $a=0.5$,
$D=0.01$ in Fig.  \ref{fig:px}), both solutions tend to converge.  Apart
from these differences, the emergence of the stable power-law
distribution, eqn.  (\ref{eq:powerlaw}), for large $x$ was already
discussed in \citep{solomon01NonStationary, solomon02, richmond01EPJB, richmond01IJMPC,
  solomon01PhysicaA, Blank-Solomon00, Sornette98, Sornette-Cont97,
  Levy-Solomon96}

\subsection{Scaling of the most probable value}
\label{sec:scaling}

In the following, we are less interested in the power-law behavior of
eqn. (\ref{eq:psx}), but mainly in the most probable value of the
process, $x_{\mathrm{mp}}$, which correspond to the peaks of the
distribution shown in Figs.  \ref{fig:x1}. From the extremum condition
\begin{equation}
  \label{eq:extrem}
0\stackrel{!}{=}\partial_xP_s(x)
\end{equation}
we find with eqn. (\ref{eq:psx}) for the most probable value:
\begin{equation}
x_{\mathrm{mp}}=\frac{a}{D-\mean{\log\lambda}} 
\label{eq:xmpsol}
\end{equation}
(Note again the difference to the result $x_{\mathrm{mp}}=a/(D+a)$
obtained in \citep{malcai}  -- for a different stochastic process).

In order to derive the scaling of $x_{\mathrm{mp}}$ on the parameters $a$
and $q_{0}$, we have to calculate the diffusion constant $D$, eqn.
(\ref{eq:diff}), and, hence, have to specify the stochastic process,
$\lambda(t)$, eqn. (\ref{eq:lambda_budget}), or $r(t)$ respectively. From
eqs. (\ref{eq:diff}), (\ref{eq:lambda_budget}) we find with $q(t)=q_{0}$
\begin{equation}
\label{eq:dapprox}
D= \mean{\log^2(1+q_0\,r)}-\mean{\log(1+q_0\,r)}^2 
\approx q_0^2\left(\mean{ r^2}-\mean{ r}^2\right)
\end{equation}
Because of $\mean{ r}=0$, this results in first approximation in the
following expression for the most probable value:
\begin{equation}
  \label{eq:xmpappoxsol}
  x_{\mathrm{mp}}\;\approx\;\frac{a}{q_0^2\,\mean{ r^2}}.  
\end{equation}
Note that this yields a good approximation only for $\mean{log
  \lambda}\approx 0$ and small values of $q_{0}$.

For the specification of $r(t)$ we refer to the four different
distributions listed in Sect. \ref{sec:InvestmentModel} and already used
in the computer simulations of Sect. \ref{sec:Simulations}. We find the
following results: 
\begin{enumerate}
\item If the returns, $r(t)$, are randomly drawn from the binomial
  distribution $B\{-1;1\}$, $\mean{r^2}=1$ holds and the scaling is
  obtained as:
\begin{equation}
x_{\mathrm{mp}}=\frac{a}{q_0^2} 
\label{eq:xmpbinomial}
\end{equation}

\item If the returns are randomly drawn from the uniform distribution
$U(-1,1)$, we find 
\begin{equation}
  \label{eq:r2uni}
\mean{r^2}=\frac{1}{2}\int_{-1}^1r^2\,dr =\frac{1}{3}
\end{equation}
and for the scaling 
\begin{equation}
x_{\mathrm{mp}}=3\frac{a}{q_0^2} 
\label{eq:xmpuniform}
\end{equation}

\item If the returns are randomly drawn from the Gaussian distribution
$N(0,0.1)$, the deviation is given by $\mean{ r^2}=\sigma^2 = 0.01$
and the scaling is obtained as
\begin{equation}
  x_{\mathrm{mp}}= 100\frac{a}{q_0^2}   
\label{eq:xmpgaussian0_1}
\end{equation}

\item If the returns are randomly drawn from the ARCH(1) process, we find
  with $\alpha_{0}=\alpha_{1}=0.1$ \citep[p. 78]{Mantegna-Stanley-2000}
\begin{equation}
  \label{eq:r2arch}
\mean{r^2}=\sigma^{2}=\frac{\alpha_{0}}{1-\alpha_{1}}=0.111
\end{equation}
which leads to a scaling of
\begin{equation}
  x_{\mathrm{mp}}= 9\frac{a}{q_0^2}   
\label{eq:xmparch}
\end{equation}
\end{enumerate}
We note that the nature of the stochastic process in all four cases is
reflected only in the different prefactors $c$, while the scaling
function $x_{\mathrm{mp}}=c\, a/q_{0}^{2}$ remains the same.  Comparing
the analytical results with the computer simulations presented in Sect.
\ref{sec:Simulations} one can see that both the scaling function and the
values of the prefactors $c$ are in perfect agreement. This shows indeed
that the approximations made for the analytical treatment were
appropriate.

\section{Conclusions}
\label{sec:ConclusionsAndFutureWork}

In this paper, we have derived an analytical expression for the
stationary probability distribution of an investor's budget $x$, eqn.
(\ref{eq:psx}), when investing in a random environment. The nature of the
underlying stochastic process was considered in four different
distributions for the return on investment (RoI), $r(t)$. Assuming that
in every time step the investor invest a constant portion $q_{0}$ of his
current budget on which he receives an RoI and further receives a very
small amount $a$ as a constant income, we have shown that the most
probable value of the investor's budget scales with the forementioned
variables as $x_{\mathrm{mp}}=a/(q_{0}\mean{r^{2}})$, eqn.
(\ref{eq:xmpappoxsol}). This result was confirmed both by analytical
investigations and extensive computer simulations for the four different
stochastic processes.

At the end, we want to comment on the range of validity for the scaling
obtained. One of the underlying assumptions of our model was a constraint
of the RoI to values between $(-1,+1)$, where -1 means a complete loss of
the investment and thus is a reasonable lower bound, whereas +1 means a
doubling of the investment, chosen for reasons of better tractability
(i.e. $\mean{r}=0$ holds, for example). These constraints for $r(t)$ used
during the computer simulations may indeed result in deviations from the
theoretical prediction, eqn. (\ref{eq:xmpappoxsol}), if the underlying
stochastic process for $r(t)$ frequently gives values outside the
interval $(-1,+1)$, which then need to be discarded. This can be the case
both for the normal and for the ARCH distribution.

For our computer simulations, we have thus chosen small values of
$\sigma^{2}$ and $a_{0}$, $\alpha_{1}$, respectively, to control the
width of the distribution and the ``outliers''. In fact, the larger these
values, the less is the agreement between the ``truncated'' computer
simulations and the theoretical approximation based on the full range of
$r$ values. 

Nevertheless, this argument does not restrict the value of the scaling
obtained in eqn. (\ref{eq:xmpappoxsol}), which is still valid. In order
to deal with broader distributions for the RoI, one has two options:
\begin{itemize}
\item[(i)] The computer simulations can be repeated on a larger interval
  for $r(t)\in (-C,+C)$ which controls the number of ``outliers''. This
  will however not change the principal insights derived in this paper.
\item[(ii)] The analytical prediction can be improved by dealing with
  \emph{truncated} distributions. This basically affects the calculation
  of $\mean{r^{2}}$. Assuming the constraint $r(t)\in (-1,+1)$, one finds
  for example for the truncated normal distribution \citep{Robert95}
  \begin{equation}
    \label{eq:trunc}
\mean{r^{2}}=\sigma^{2} - \sigma \sqrt{\frac{2}{\pi}} \;
\frac{\exp\left\{-\frac{1}{2 \sigma^{2}}\right\}}{\operatorname{erf}\left\{
    \frac{1}{\sqrt{2}\sigma}\right\}} 
  \end{equation}
which holds also for larger $\sigma^{2}$. For truncated ARCH or GARCH
distributions, the situations are more complicated, here we refer to the
literature \citep{Engle82, Bollerslev86, Nelson92, Bera-Higgins93} 
\end{itemize}

Eventually, we wish to comment on the investor dynamics proposed in this
paper. These were by purpose related to multiplicative  stochastic
processes, to allow for analytical insights and conclusions about the
influence of different stochastic distributions. Despite its simplicity,
the dynamics of eqn. (\ref{eq:budgetmap}) is not restricted to artificial
scenarios. In fact, the dynamics of the RoI, $r(t)$, can be taken from
real time series, instead of being modeled by a stochastic process. The
most challenging application, however, is in the dynamics of the variable
$q(t)$ which decides about the portion of the budget to be invested.
While risk averse agents may tend to lower $q$, risk seeking strategies
may go for higher $q$ and for higher yields in the lucky case.  Throughout
this paper, $q$ was set to a constant but small value, independent of 
individual characteristics.  Any realistic investment scenario deals with
the question to adjust the risk propensity $q(t)$ in time based on the
observation of previous $r(t)$. For the model given, an artificial
intelligence approach to this question is presented in \citep{Navarro}.
As a next step, the investment dynamics can be extended towards a
portfolio scenario, where both $r$ and $q$ become multidimensional
variables, to allow different investment strategies for different assets.

\subsection*{Acknowledgments}

The authors gratefully acknowledge discussions with Robert Mach and  
Frank E. Walter (Zurich). We are further indepted to Jan Lorenz (Zurich)
for corrections. 

\bibliography{investor_arxiv}

\section*{Appendix}

Instead of dealing with the probability distribution $P(x)$ as in Sect.
\ref{sec:analytical}, one can try to treat the stochastic dynamics of
eqn. (\ref{eq:basic2}) directly. For this case, we can present at least a
formal solution using the $\mathcal{Z}$-transform \citep{jury73}.
Rewriting eqn. (\ref{eq:basic2}) as
\begin{equation}
  \label{eq:ay1}
  a\;=\;x(t+1)-\lambda x(t)
\end{equation}
the $\mathcal{Z}$-transform leads to
\begin{equation}
  \label{eq:trans1}
  a\sum_{n=0}^{\infty}\frac{1}{z^n}=a\,\frac{z}{z-1}= 
z\left[X(z)-x(0)\right]-\lambda X(z)
\end{equation}
where $X(z)$ is given by
\begin{equation}
  \label{eq:trans2}
  X(z)=x(0)\,\frac{z}{z-\lambda}+a\,\frac{z}{(z-1)(z-\lambda)}
\end{equation}
Using the inverse tranform 
\begin{equation}
\label{eq:transinv}
X(z^{-1})
=\left[x(0)-\frac{a}{1-\lambda}\right]\,\frac{1}{1-\lambda z}
+\frac{a}{1-\lambda}\,\frac{1}{1-z}
\end{equation}
the solution for $x(t)$ can be found as
\begin{eqnarray}
\label{eq:solve}
x(t)&=&\frac{1}{t!}\,\partial_z^t\,X\left(\frac{1}{z}\right)\\
&=&\lambda^t\,\left[x(0)-\frac{a}{1-\lambda}\right]
+\frac{a}{1-\lambda}\\
&=&\lambda^t\,x(0)+a\,\frac{1-\lambda^t}{1-\lambda} \\
&=&\lambda^t\,x(0)+a\,\sum_{s=0}^{t-1}\lambda^s
\end{eqnarray}
From this solution, we see that the decisive condition on $\lambda$ for a
well-defined solution reads:
\begin{equation}
  \label{eq:lambda1}
  |\lambda|<1
\end{equation}
which agrees with the finding of $\mean{\log|\lambda|}<0$  obtained from
the treatment of the probability distribution $P(x)$.

\end{document}